%% file: TMR-Magnetite.tex
\documentclass[aip,graphicx]{revtex4-1}
\usepackage{graphicx} 
\usepackage{multirow}
\usepackage{xcolor}
\renewcommand{\figurename}{Figure}

\draft % marks overfull lines with a black rule on the right

\begin{document}

% Use the \preprint command to place your local institutional report number 
% on the title page in preprint mode.
% Multiple \preprint commands are allowed.
%\preprint{}
\title{Annealing Effects on the Magnetic and Magnetotransport Properties of Iron Oxide Nanoparticles Self-Assemblies} 
%Title of paper

% repeat the \author .. \affiliation  etc. as needed
% \email, \thanks, \homepage, \altaffiliation all apply to the current author.
% Explanatory text should go in the []'s, 
% actual e-mail address or url should go in the {}'s for \email and \homepage.
% Please use the appropriate macro for the type of information

% \affiliation command applies to all authors since the last \affiliation command. 
% The \affiliation command should follow the other information.

\author{Fernando Fabris}
\email[]{ffabris@ifi.unicamp.br}
%\homepage[]{Your web page}
%\thanks{}
%\altaffiliation{}
\affiliation{Instituto de Física “Gleb Wataghin,” UNICAMP, 13083-859 Campinas, São Paulo, Brazil}
\affiliation{Resonancias Magnéticas, Gerencia de Física, Centro Atómico Bariloche, Av. Bustillo 9500, (8400) S. C. de Bariloche (RN), Argentina.}

\author{Enio Lima Jr.}
\affiliation{Resonancias Magnéticas, Gerencia de Física, Centro Atómico Bariloche, Av. Bustillo 9500, (8400) S. C. de Bariloche (RN), Argentina.}
\affiliation{Instituto de Nanociencia y Nanotecnología (CNEA-CONICET), Nodo Bariloche, Av. Bustillo 9500, (8400) S. C. de Bariloche (RN), Argentina.}

\author{Jorge Martín Nuñez}
\affiliation{Resonancias Magnéticas, Gerencia de Física, Centro Atómico Bariloche, Av. Bustillo 9500, (8400) S. C. de Bariloche (RN), Argentina.}
\affiliation{Instituto de Nanociencia y Nanotecnología (CNEA-CONICET), Nodo Bariloche, Av. Bustillo 9500, (8400) S. C. de Bariloche (RN), Argentina.}
\affiliation{Instituto Balseiro, CNEA-UNCuyo, Av. Bustillo 9500, (8400) S. C. de Bariloche (RN), Argentina}
\affiliation{Instituto  de  Nanociencias  y  Materiales  de  Aragón, CSIC-Universidad  de  Zaragoza Mariano Esquillor s/n, Zaragoza, E-50018, Spain}

\author{Horacio E. Troiani}
\affiliation{Caracterización de Materiales,Centro Atómico Bariloche, Av. Bustillo 9500, (8400) S. C. de Bariloche (RN), Argentina.}

\author{Myriam H Aguirre}
\affiliation{Instituto  de  Nanociencias  y  Materiales  de  Aragón, CSIC-Universidad  de  Zaragoza Mariano Esquillor s/n, Zaragoza, E-50018, Spain}
\affiliation{Dept. Física de la Materia Condensada, Universidad de Zaragoza, C/ Pedro Cerbuna 12, 50009, Zaragoza, Spain}
\affiliation{Laboratorio de Microscopías Avanzadas, Universidad de Zaragoza, Mariano
Esquillor s/n, 50018, Zaragoza, Spain}

\author{Victor Leborán}
\affiliation{Centro Singular de Investigación en Química Biolóxica e Materiais Moleculares (CIQUS), Departamento de
Química-Física, Universidade de Santiago de Compostela, Santiago de Compostela 15782, Spain}

\author{Francisco Rivadulla}
\affiliation{Centro Singular de Investigación en Química Biolóxica e Materiais Moleculares (CIQUS), Departamento de
Química-Física, Universidade de Santiago de Compostela, Santiago de Compostela 15782, Spain}

\author{Elin L. Winkler}
\affiliation{Resonancias Magnéticas, Gerencia de Física, Centro Atómico Bariloche, Av. Bustillo 9500, (8400) S. C. de Bariloche (RN), Argentina.}
\affiliation{Instituto de Nanociencia y Nanotecnología (CNEA-CONICET), Nodo Bariloche, Av. Bustillo 9500, (8400) S. C. de Bariloche (RN), Argentina.}
\affiliation{Instituto Balseiro, CNEA-UNCuyo, Av. Bustillo 9500, (8400) S. C. de Bariloche (RN), Argentina}

%\homepage[]{Your web page}
%\thanks{}
%\altaffiliation{}

% Collaboration name, if desired (requires use of superscriptaddress option in \documentclass). 
% \noaffiliation is required (may also be used with the \author command).
%\collaboration{}
%\noaffiliation

\date{\today}

\begin{abstract}
In magnetic tunnel junctions based on iron oxide nanoparticles the disorder and the oxidation state of the surface spin as well as the nanoparticles functionalization play a crucial role in the magnetotransport properties. In this work, we report a systematic study of the effects of vacuum annealing on the structural, magnetic and transport properties of self-assembled $\sim$10 nm Fe$ _3$O$_4$ nanoparticles. The high temperature treatment (from 573 K to 873 K) decomposes the organic coating into amorphous carbon, reducing the electrical resistivity of the assemblies by 4 orders of magnitude. At the same time, the 3.Fe$^{2+}$/(Fe$^{3+}$+Fe$^{2+}$) ratio is reduced from 1.11 to 0.13 when the annealing temperature of the sample increases from 573 K to 873 K, indicating an important surface oxidation. Although the $~2$ nm physical gap remains unchanged with the thermal treatment, a monotonous decrease of tunnel barrier width was obtained from the electron transport measurements  when the annealing temperature increases, indicating an increment in the number of defects and hot-spots in the gap between the nanoparticles. This is reflected in the reduction of the spin dependent tunneling, which reduces the interparticle magnetoresistance. This work shows new insights about influence of the nanoparticle interfacial composition, as well their the spatial arrangement, on the tunnel transport of self-assemblies, and evidence the importance of optimizing the nanostructure fabrication for increasing the tunneling current without degrading the spin polarized current.

\end{abstract}

% insert suggested PACS numbers in braces on next line

\maketitle

%\maketitle must follow title, authors, abstract and \pacs

% Body of paper goes here. Use proper sectioning commands. 
% References should be done using the \cite, \ref, and \label commands
\section{Introduction}

 The spin- dependent transport has been extensively studied in different nanostructures for exploiting the possibilities of controlling spin injection. Since the discovery of giant magnetoresistance \cite{Baibich1988, Binasch1989} new nanotechnologies have been developed based on magnetic tunnel junction as magnetic recording \cite{Hu2012, Ohkoshi2020}, spin valves \cite{Elahi_2022} and spintronic-based logic \cite{Black2000, Shen1997, Ahn2020}. A magnetic tunnel junction is a trilayer structure in which two ferromagnetic (FM) metallic layers are separated by a thin insulator, \cite{Julliere1975,Moodera1995} so that the tunneling transport is controlled by the relative orientation of the magnetization of the FM.  Similar behavior was also observed in granular thin films,\cite{Suchaneck2022} such as CrO$_2$\cite{Coey1998}, Fe$_3$O$_4$\cite{Taub2009}, La$_{0.7}$Sr$_{0.3}$MnO$_3$\cite{Li2005}, La$_{0.67}$Ca$_{0.33}$MnO$_3$ \cite{Wagenknecht2006,Rivas2000}, where the grain boundaries act as tunnel barrier. Among these compounds, magnetite (Fe$_3$O$_4$) is one of the more attractive materials due to its half-metallic properties, high spin polarization at room temperature\cite{Wang2006,Mitra2018,Zhang2022}, natural abundance, low cost and being non-toxic for the environment. In the last years important advances have been made in the fabrication of Fe$_3$O$_4$ based nanostructures for studying the tunnel magnetoresistance, \cite{Wang2023} magnetoelectric coupling \cite{Dong2022} and the design of spintronic devices \cite{Ansari2021, ANSARI2020}. Noncrystalline iron oxide thin films and hybrid/composites were fabricated from different methods as chemical vapor deposition (CVD), pulsed laser deposition (PLD) or sputtering. On the other hand, chemicals routes as the thermal decomposition of organometallic precursors in the presence of surfactants or the  polyol method, offer the possibility of synthesizing nanoparticles with controlled chemical composition, shape, and narrow size distribution; which can be self-assembled over large areas. \cite{Wang2009,Kohiki2013,Mitra2016,Mitra2018, Dong2010,Zhou2019, Mi2018, Yang2023, Fabris2019PRA}  In these structures the nanoparticles organic coating, resulting from the synthesis, acts as a tunneling barrier that controls the electronic transport. In particular, these chemical methods are interesting for the design of flexible spintronic devices, as the nanoparticles can be deposited onto the plastic flexible substrate at room temperature,\cite{Kurokawa2022} and also for the design of spin valves. \cite{Zhou2019, Chen2013} On the other hand, the organic capping ligands that coat the nanoparticles, prevent their agglomeration and promote the order of the assembly in the final device; but also constitute a very insulating barrier which impedes interparticle electronic transport. Therefore, to increase the electrical conductivity of the assembly, a subsequent annealing at higher temperature is a common procedure. \cite{Chen2013,Jiang2017,Fabris2019PRA, Jang2007, Kurokawa2022} In most experimental observations of epitaxial or nanocrystalline Fe$_3$O$_4$ thin films, the tunnel magnetoresistance (TMR) value  is much lower than the theoretically predicted for half-metallic compounds, where a fully spin polarized, P = 100\%,  is expected. This reduction is ascribed to the diminution of spin polarization due to surface oxidation, surface spin disorder, and/or electron-electron correlations. \cite{Huang2002, Morton2002} These effects can be exacerbated by the thermal annealing, which is usually performed in vacuum or inert atmosphere.  Therefore, additional Fe$_3$O$_4$ functionalization with organic molecules as acetic acid, amine, polyvinyl acetate,\cite{Wang2009, Yue2011, Mitra2016, Roy2022, Roma2023}  or inorganic coatings as SiO$_2$ or ZrO$_2$ \cite{Zeng2006, Wang2007, Arnay2021, Mi2018} were studied in order to enhance the room-temperature magnetoresistance, by improving the NPs arrays and/or protecting the nanoparticle surface from oxidation. However, this step complicates the fabrication process and it is found that the enhancement of TMR is still far away from the predicted value.

In this context, we are interested in finding the conditions that optimize and simplify the fabrication of nanoparticles assemblies in order to enhance the electrical conductivity of the films preserving their spin polarization. With this aim we report a detailed study of the effects of vacuum thermal annealing on the structural, magnetic and magnetotransport properties of self-assemblies of $\sim$10 nm Fe$_3$O$_4$  nanoparticles coated with oleic acid. The structural study shows that the size of nanoparticles and their arrangement do not change appreciably with the annealing up to 873 K. However, a substantial  increase of surface Fe$^{3+}$/Fe$^{2+}$-ratio occurs, due to the oxidation of the nanoparticle surface by the decomposition of the oleic acid. Although the partial elimination of the organic barrier reduces the electrical resistance of the system, the surface oxidation decreases the spin polarization and partially suppresses  the TMR. Therefore, we show that the annealing conditions must be carefully optimized for the fabrication of functional nanoparticle self-assemblies, in order to obtain an efficient control on the tunneling current and the spin polarization by means of the spatial arrangement and the surface composition.
 
\section{Experimental}
The Fe$_3$O$_4$ nanoparticles were synthesized by thermal decomposition of organo-metallic precursors at high temperature following the procedure described in previous references. \cite{Sun2002, Fabris2019} Briefly, 12 mMol of Fe(III) acetylacetonate were used in presence of 23 mMol of 1,2-octanediol, 12 mMol of Oleic acid, 29 mMol of Oleylamine and 210 mMol of Benzyl ether as solvent. This solution is heated at 473 K, kept for 20 minutes under N$_2$ flow (0.1 mL/min) and intense mechanical stirring, followed by heating at 15 K/min up to the reflux temperature (563 K) and keeping in it during one hour. After cooling the sample to room temperature, it was washed by
adding ethanol and acetone in the proportion 10:5 V/V and magnetically separated. Finally, the NPs are dispersed in hexane with a concentration of 5 mg/ml. 

The self-assembly of Fe$_3$O$_4$ nanoparticles was made by the liquid-air interface method  following the procedure reported in Refs. \cite{Fabris2019PRA,Jiang2017,Chen2013,Dong2010}, and schemitized in Fig. \ref{Esquema}. In brief, 10 $\mu$l of the solution of nanoparticles in hexane was drop casted onto the surface of triethylene glycol in a $1.5 \times 1.5 \times 1.0$ cm$^3$ Teflon container.  A self-assembled film is formed after complete evaporation of hexane (after 10-15 min). After that, the triethylene glycol is removed very slowly using a syringe to gently deposit the assembled film on the glass substrate strategically positioned within the Teflon container. Finally, the self-assemblies were annealed at a constant temperature in a range between 573 K and 873 K by 30 minutes under mechanical vacuum ($\approx 10^{-3}$ Torr). The samples were labeled as SX, where X corresponds to the annealing temperature in Kelvin.  

Transmission electron microscopy (TEM) images were taken in a Philips CM200 microscope equipped with an Ultra-Twin lens operating at 200 kV and in a field emission FEI Tecnai G2 F30 microscope operating at an accelerating voltage of 300 kV.  In order to perform the TEM analysis the  Fe$_3$O$_4$ self-assemblies were transferred from the triethylene glycol surface to commercial silicon nitride TEM grids followed by the corresponding thermal annealing. The diameter distribution $f(D)$ was obtained by measuring $\sim400$ particles assuming spherical shape. The histograms were fitted with a lognormal function $f(D)=(1/\sqrt{2\pi}\sigma_0 D)exp[-ln^2(D/D_0)/2\sigma_0^2]$, from which mean diameter $\left\langle D\right\rangle$ and standard deviation $\sigma$ was obtained.

X-ray photoelectron spectroscopy (XPS) was performed with a hemispherical electrostatic energy analyzer (r = 10 cm) using Al $K_\alpha$ radiation (h$\nu$ = 1486.6 eV). The binding-energy (BE) scale was calibrated with the position of the C1s peak corresponding to the residual carbon coating or adsorbed carbon.

Magnetic measurements of the self-assemblies were performed in a SQUID magnetometer (Quantum Design, MPMS3) with the magnetic field applied in the plane of the substrate. The magnetization as a function of temperature (M(T)) was measured with an applied field of 100 Oe by using the Zero-field-cooling (ZFC) and Field-cooling (FC) protocols. 

The magnetotransport measurements were performed using a Keithley 4200 source-measure unit in a two-probe configuration, with a maximum applied field of $\pm12$ kOe. For these measurements the self-assembled nanoparticles were deposited on commercial substrates of Si/SiO$_2$ (200 nm thick) patterned with interdigitated Au electrodes (separated from 2.5 to 10 $\mu$m and total effective length from 6 to 10 mm) followed by the thermal annealing.
\section{Results and discussions}

\begin{figure}
	\centering
	\includegraphics[width=0.8\textwidth]{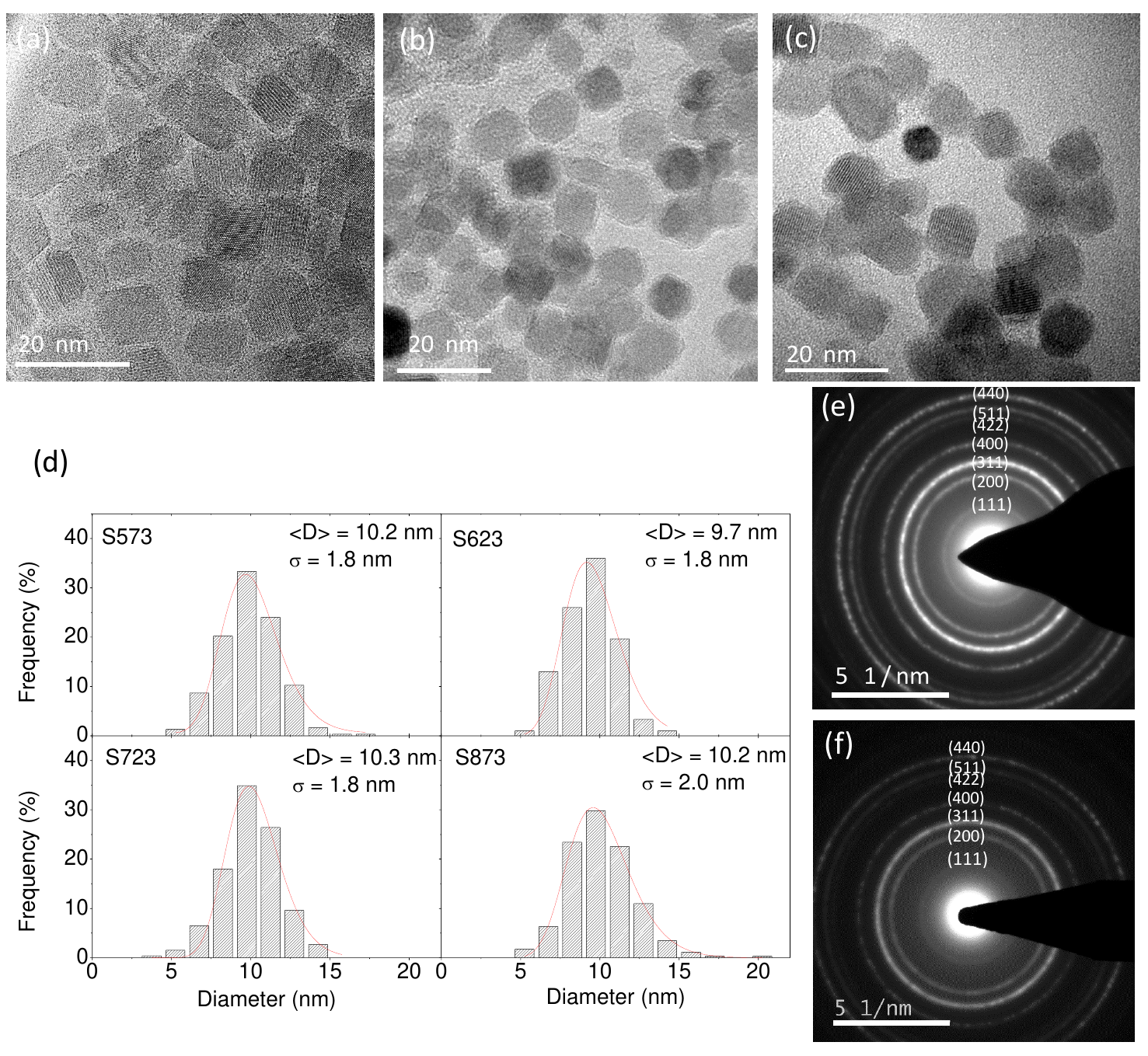}
      \caption{TEM images of the samples S573 (a), S723 (b) and S873 (c) and the corresponding size distribution are shown in (d), where the red line is the fit with a lognormal distribution. Electron diffraction of the sample S573 (e) and S873 (f), where the spinel  diffraction rings are indicated.}
      \label{TEM}
\end{figure}

Figure \ref{Esquema} shows a representative scheme of the sample preparation and figure \ref{XRD} of Supplementary Material shows the X- ray diffraction pattern of the Fe$_3$O$_4$ as-synthesized nanoparticles where the diffraction peaks are indexed with the Fd3m  space group of the Fe$_3$O$_4$ spinel phase. Figures \ref{TEM} (a), (b) and (c) show high resolution TEM images of the Fe$_3$O$_4$ nanoparticles annealed at 573 K, 723 K and 873 K, respectively. Notice that from this annealing condition, it is not generated significant size or morphological changes in the particles, see figure \ref{TEM} (d); although for the highest temperature, i.e 873K, we observed that some NPs start to neck together. We ascribed this behavior to the presence of residual carbon that remains after the annealing and prevents NPs from coalescing up to 873 K.\cite{Fabris2019PRA} Also the self-assemblies form a stable superstructure that preserves the gap between the NPs under the annealing, which also prevents them from coalescing. 

The electron diffraction pattern of the sample S573 and S873 shown in the figure \ref{TEM} (e) and (f), respectively, is characteristic of the Fd-3m space group of magnetite with a lattice parameter of 8.41$\pm$0.02 \(\text{\AA}\). The samples annealed at lower temperatures show similar electron diffraction patterns; and no extra iron oxide phases were observed in the high resolution images or electron diffraction measurement.

\begin{figure}
	\centering
	\includegraphics[width=1\textwidth]{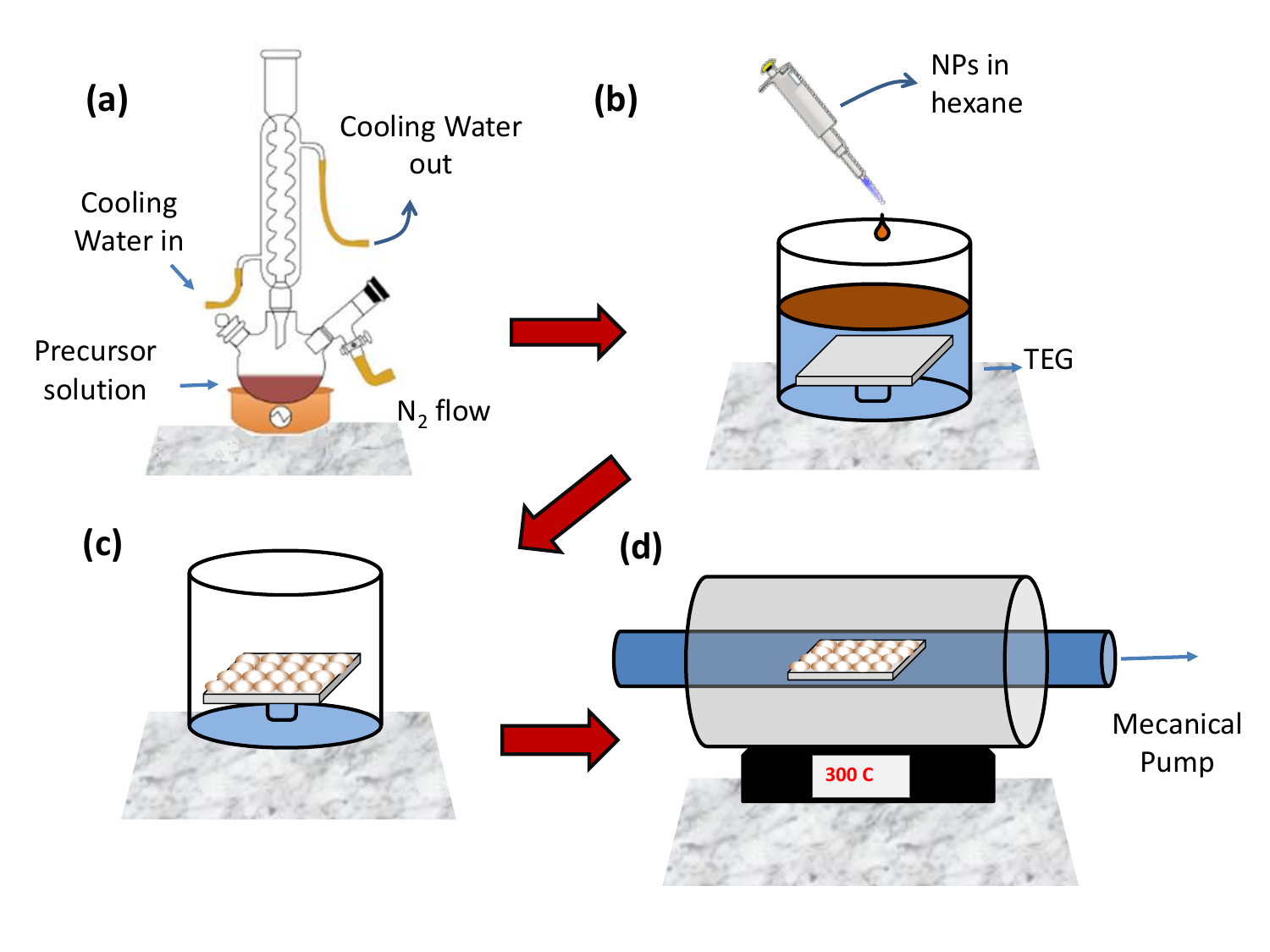}
      \caption{Schematic representation of the assembly preparation (a) First the nanoparticles were synthesized by thermal decomposition method. The obtained solution was washed several times and dispersed in hexane at a 5 mg/ml concentration. (b) Then, the self-assembled was formed by drop casting the nanoparticles hexane solution onto the surface of triethylene glycol (TEG),(c)  The self-assembled nanoparticles were deposited onto a glass substrate by slowly removing the TEG. (d) Finally, the assembly was annealed in mechanical vacuum at the desired temperature. }
      \label{Esquema}
\end{figure}

\begin{figure*}
	\centering
	\includegraphics[width=0.7\textwidth]{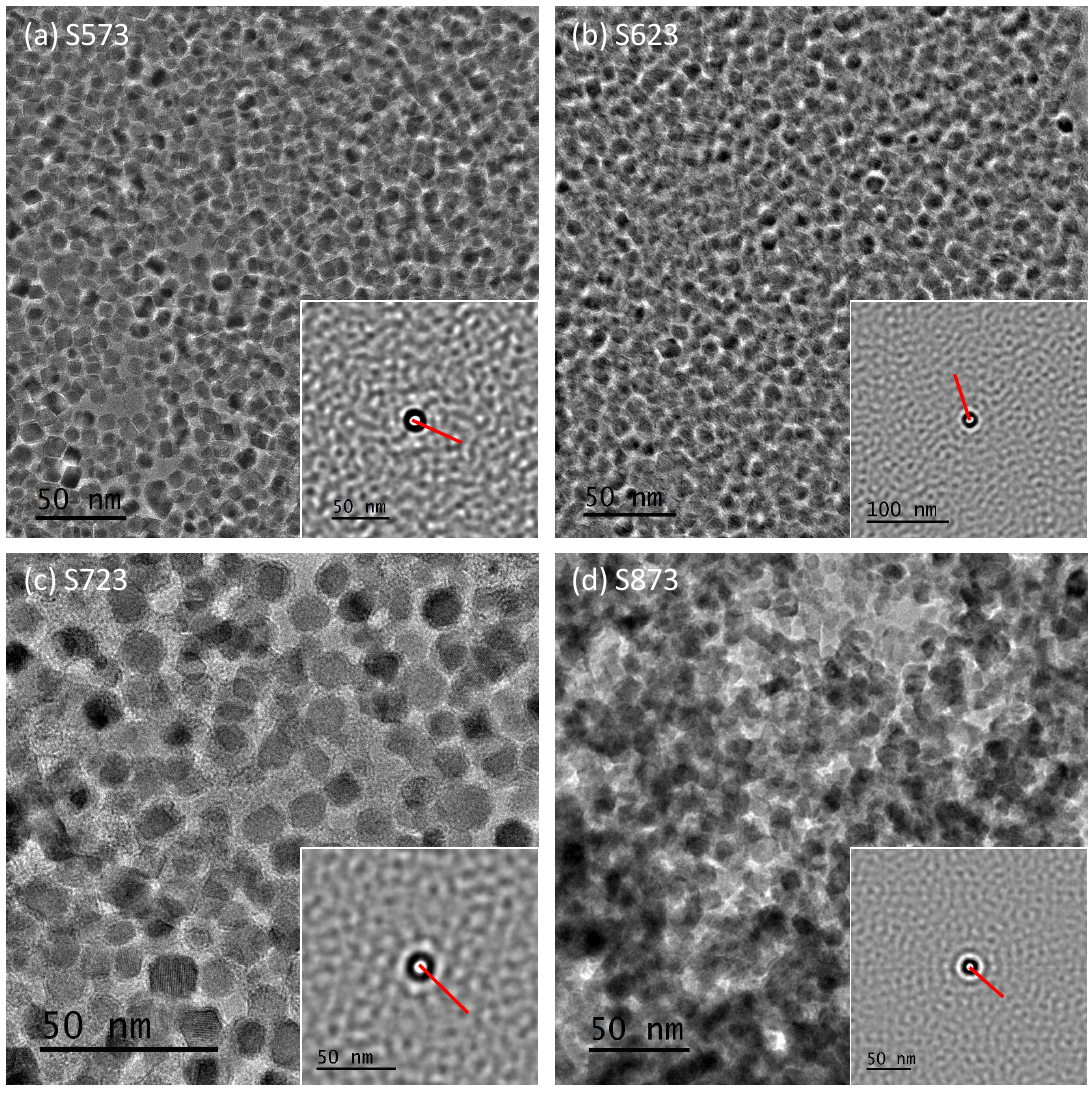}
	\caption{Low magnification TEM images of the self-assemblies annealed at 573 K (a), 625 K (b), 723 K (c) and 873 K (d). The inset shows the auto-correlation image of each figure. The red line indicates the position of intensity analysis.}
	\label{TEM-LM}
\end{figure*}

Figure \ref{TEM-LM} presents low magnification TEM images of the self-assembled films deposited on  TEM grids followed by the corresponding thermal annealing, where large areas of self-assembled nanoparticles can be observed. The self-assembled film maintains its main features as the temperature of the annealing increases until at 873 K, where some particles start to neck together.\cite{Fabris2019PRA} To obtain information of the film arrangement and degree of ordering the autocorrelation spectra have been calculated from the TEM images (see the insets of Figure \ref{TEM-LM}). From the patterns, one may conclude that the particulate films have an approximately local hexagonal order.\cite{Pauly2011} The radial profiles of the correlation image can be described by $C(r)=A exp\left( \frac{-r}{\xi}\right)  sin\left(\pi\left(\frac{r-r_c}{\omega}\right) \right)$, where $r$ is the distance measured on the autocorrelation profile, $r_c$ the phase difference, $A$ the amplitude, $\xi$ the correlation length and $\omega=a/2$ is related to the center to center particles distance $a$.\cite{Pauly2011,Wu2008,Pichler2011} The intensity profile, taken along the red line of the inset of Figure \ref{TEM-LM}, with the corresponding fitting curve is shown in the figure \ref{Correlation} of the Supplementary Material. The obtained $a$ parameter is $\sim 12 nm$ for all the annealing temperatures, although the obtained $\xi$ changes from 5 nm to 10 nm, which indicates a poor order in the film arrangement. From the $a$ value and the average NPs size, a gap between the NPs of $\sim 2$ nm can be estimated. This value is in agreement with the one expected for NPs coated with oleic acid molecules (2-3 nm).\cite{Pauly2011} This result indicates that, although the amount of residual carbon decreases with the temperature \cite{Fabris2019PRA, Lavorato2021}, the gap between the nanoparticles does not change significantly. 

\begin{figure}
	\centering
	\includegraphics[width=\textwidth]{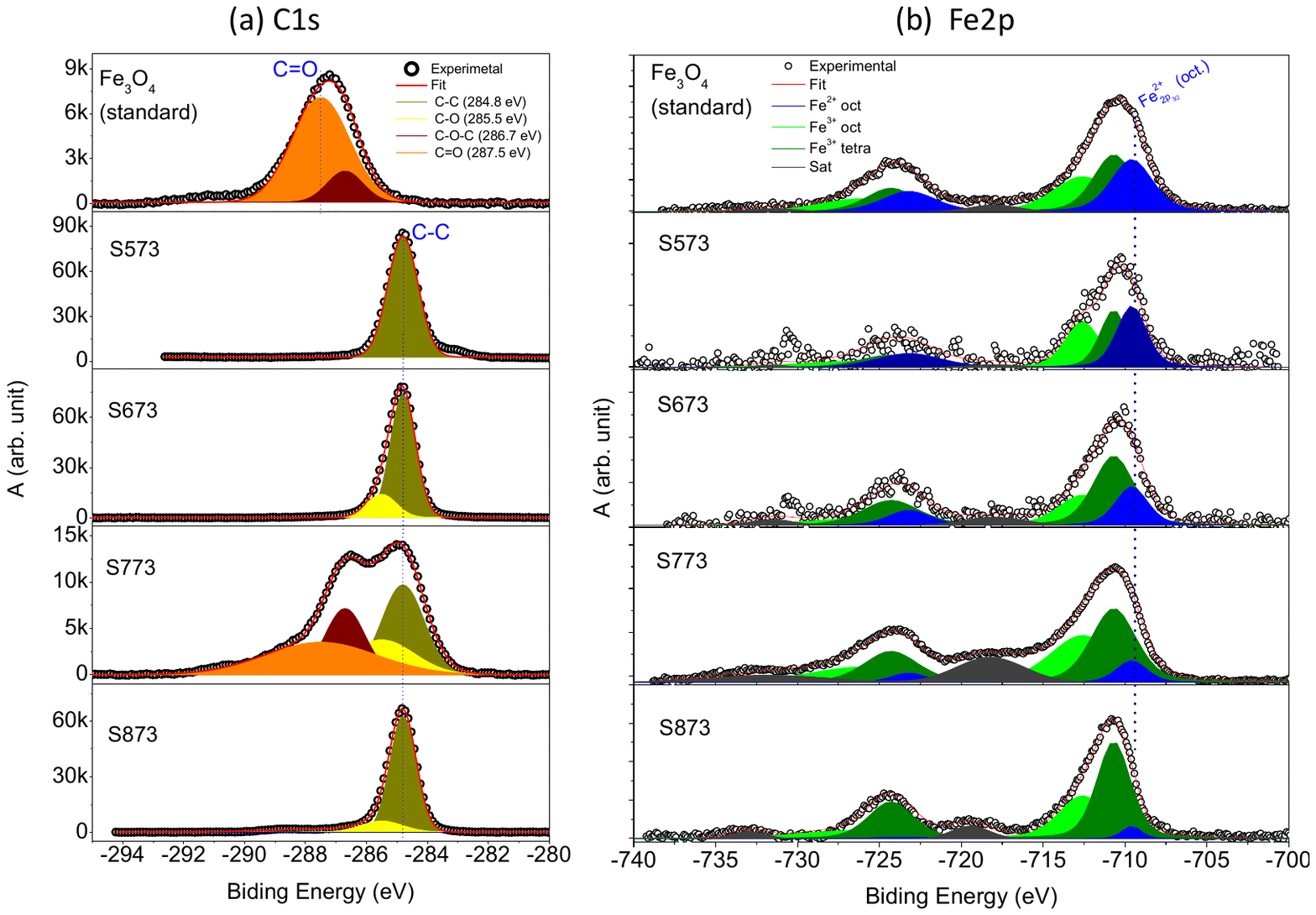}
	\caption{Background subtracted XPS spectrum in the C1s region (a) and Fe2p region (b) of Fe$_3$O$_4$ pattern sample and nanoparticles self-assemblies with the corresponding fitting curves (red line). The blue dotted vertical line marks the position of the  Fe$^{2+}_{2p_{3/2}}$ peak.}
	\label{XPS}
\end{figure}

The tunnel transport can be strongly affected by the surface composition, changing not only the height of the barrier but also the spin polarization; therefore, a careful chemical analysis of the nanoparticles surface is demanded.  Figure \ref{XPS} shows the XPS spectra of the Fe$_3$O$_4$ self-assemblies, as well as the spectrum of the bulk Fe$_3$O$_4$ as a reference. The binding-energy scale should be corrected due to the surface charging, effect that is more important in the self-assembled films due to its high electrical resistance. The standard method for this calibration is the use of the surface carbon species XPS peaks. In self-assembled samples the predominant peak corresponds to C-C (284.8 eV), while C-O (285.5 eV), C-O-C (286.7 eV) and C=O (287.5 eV) peaks are also identified in some samples but with much lower intensity;\cite{Tien2011, Sadri2017, Morais2015} instead, for the bulk magnetite sample the C=O peak located at 287.5 eV is clearly detected while the C-C peak is not resolved.  Therefore, the correction was made using the C=O peak located at 287.5 eV for the standard magnetite sample and for the Fe$_3$O$_4$ self-assemblies, we used the C-C peak at 284.8 eV. 

Figure \ref{XPS} (b) shows the XPS spectra of the standard and self-assemblies at Fe2p region. The most noticeable effect is the systematic shift to lower binding energy of the Fe2p signal with the increase of the annealing temperature, which, according to the spectrum of the Fe$^{2+}_{2p_{3/2}}$ standard, indicates an increase in the proportion of Fe$^{3+}$. To obtain  quantitative information, the XPS spectra was fitted with the peaks at -709.4, -710.5 and -712.5 eV related to Fe$^{2+}2p_{3/2}$ and Fe$^{3+}2p_{3/2}$ in the octahedral sites and Fe$^{3+}2p_{3/2}$ in the tetrahedral site.\cite{Grosvenor2004} For the fitting it was considered that the  Fe$^{3+}2p_{1/2}$ peak has a half intensity and it is separated by -13.6 eV from the Fe$2p_{3/2}$ peak. Furthermore, two satellites located at -718 eV of the Fe$^{3+}2p_{3/2}$ and -731.6 eV of the Fe$^{3+}2p_{1/2}$ were considered.  The position of each peak was fixed and only its width at half maximum ($\sigma$) and area were adjusted, using a fixed shape factor of 0.6 for all the Voigt profiles. The fitting curves are shown with red lines in the figure \ref{XPS} (b) and the adjusted parameters were summarized in the table \ref{XPS-results} of the Supplemental Material. From the Fe$^{2+}$ and Fe$^{3+}$ peaks intensities the relation $3*Fe^{2+}/(Fe^{2+}+Fe^{3+})$ were calculated. As expected, the Fe$_3$O$_4$ standard presents $3*Fe^{2+}/(Fe^{2+}+Fe^{3+})\approx1$. On the other hand, this ratio decreases with increasing the annealing temperature, from 1.11 for the sample annealed at 573K, 0.68 at 673 K, 0.29 at 773 K  and, finally, reaching a value of 0.13 at 873 K. This result confirms a partial surface oxidation from  Fe$^{2+}$Fe$^{3+}_2$O$_4$ magnetite to $\gamma$-Fe$^{3+}_2$O$_3$ maghemite or $\alpha$-Fe$^{3+}_2$O$_3$ hematite as the annealing temperature increases, despite being performed in vacuum.\cite{Jafari2015,Schwaminger2017} 

\begin{figure}
	\centering
	\includegraphics[width=0.8\textwidth]{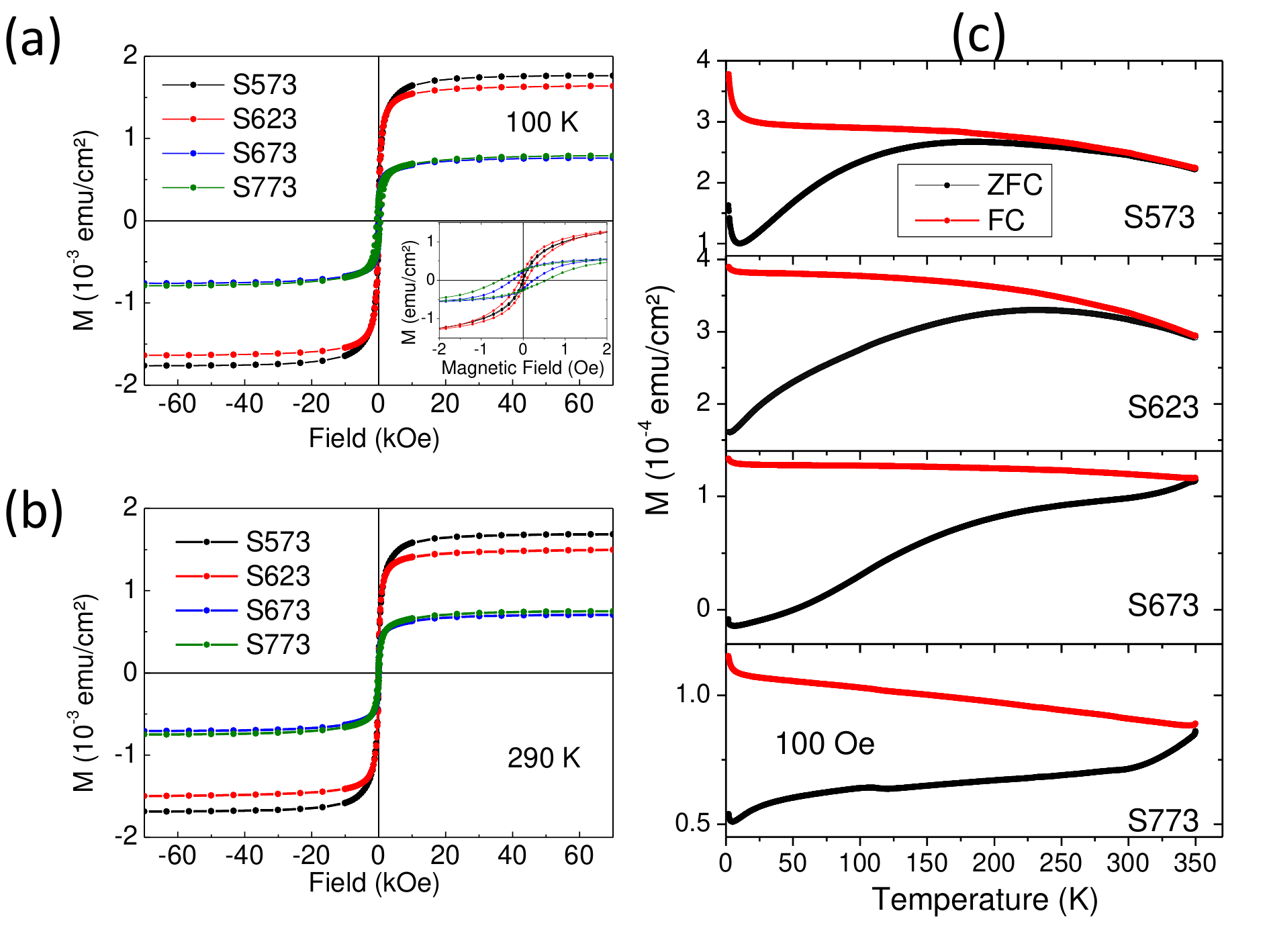}
	\caption{Magnetization as function of the field at 100 K(a) and 290 K(b) and ZFC-FC curves (c) of the samples annealed at different temperatures. Inset in (a) is a zoom of the magnetization at low field.}%
	\label{Mag}
\end{figure}

The figures \ref{Mag} (a) and (b) show the magnetization normalized by the sample film (self-assemblies) area measured at 100 K and 290 K for annealed self-assemblies. At room temperature the samples show M(H) reversible behavior, consistent with the superparamagnetism of the  assembled $\sim$ 10 nm NPs.\cite{Fabris2019PRA} At low temperature a small coercivity field is observed, increasing as the annealing temperature does, from $H_C=5$ Oe for S573 to $H_C=550$ Oe for S873. The saturation magnetization is also reduced with the annealing temperature, consistent with the oxidation of magnetite ($M_S\sim 480$ emu/cm$^3$) to maghemite ($M_S\sim 430$ emu/cm$^3$). Thermal annealing temperature above 673 K results in a large drop of the saturation magnetization, to a half of its initial value, pointing to a partial oxidation of the nanoparticles to the antiferromagnetic $\alpha$-Fe$_2$O$_3$ ($M_S\sim 3$ emu/cm$^3$). Also notice that  $M_S$ remains almost constant when the temperature diminishes, indicating that the fraction of thermally fluctuating surface spins is small  (due to the presence of interparticle interactions in the assembled NPs), in agreement wiht previous results. \cite{Winkler2004}

The irreversibility temperature of the ZFC-FC magnetization curves and the maximum of the ZFC curve, related to the mean blocking temperature ($T_B$), both increase with the annealing temperature. For magnetic monodomain nanoparticle an estimation of $T_B$ can be obtained from $T_B=ln(\tau_0/\tau_m)K_{eff}V/k_B$ where $K_{eff}$ is the effective magnetic anisotropy and $V$ the magnetic volume of the particle, $\tau_0$ and $\tau_m$ are the attempt relaxation time and the characteristic measuring time of the system, respectively. The characteristic time for SQUID measurements  is $\tau_m\sim$60 s while $\tau_0\sim$ 10$^{-10}$ s is assumed for ferrimagnetic NPs system. As the particle size and the gap between the nanoparticle do not change with annealing temperature (see Figure \ref{TEM}), the increase of $T_B$ can be attributed to an increase in the $K_{eff}$.  The $K_{eff}$ in nanoparticles systems depends of the magnetocrystaline anisotropy of the magnetic phase, surface anisotropy, shape anisotropy and also effects of  magnetic interactions between the particles. \cite{Knobel2008,Fabris2019JAP} For bulk Fe$_3$O$_4$ is $K_1=-1.35\times10^5$ erg/cm$^3$, for $\gamma$-Fe$_2$O$_3$ is $K_1=-4.65\times10^4$ erg/cm$^3$ and for $\alpha$-Fe$_2$O$_3$ is $K_1=1.2\times10^7$ erg/cm$^3$.\cite{Dunlop2007} Therefore, we attributed the increase in $K_{eff}$ to the variation of the magnetocrystalline anisotropy due to the improvement of the crystallinity and also the oxidation of the sample.

\begin{figure}
	\centering
	\includegraphics[width=0.7\textwidth]{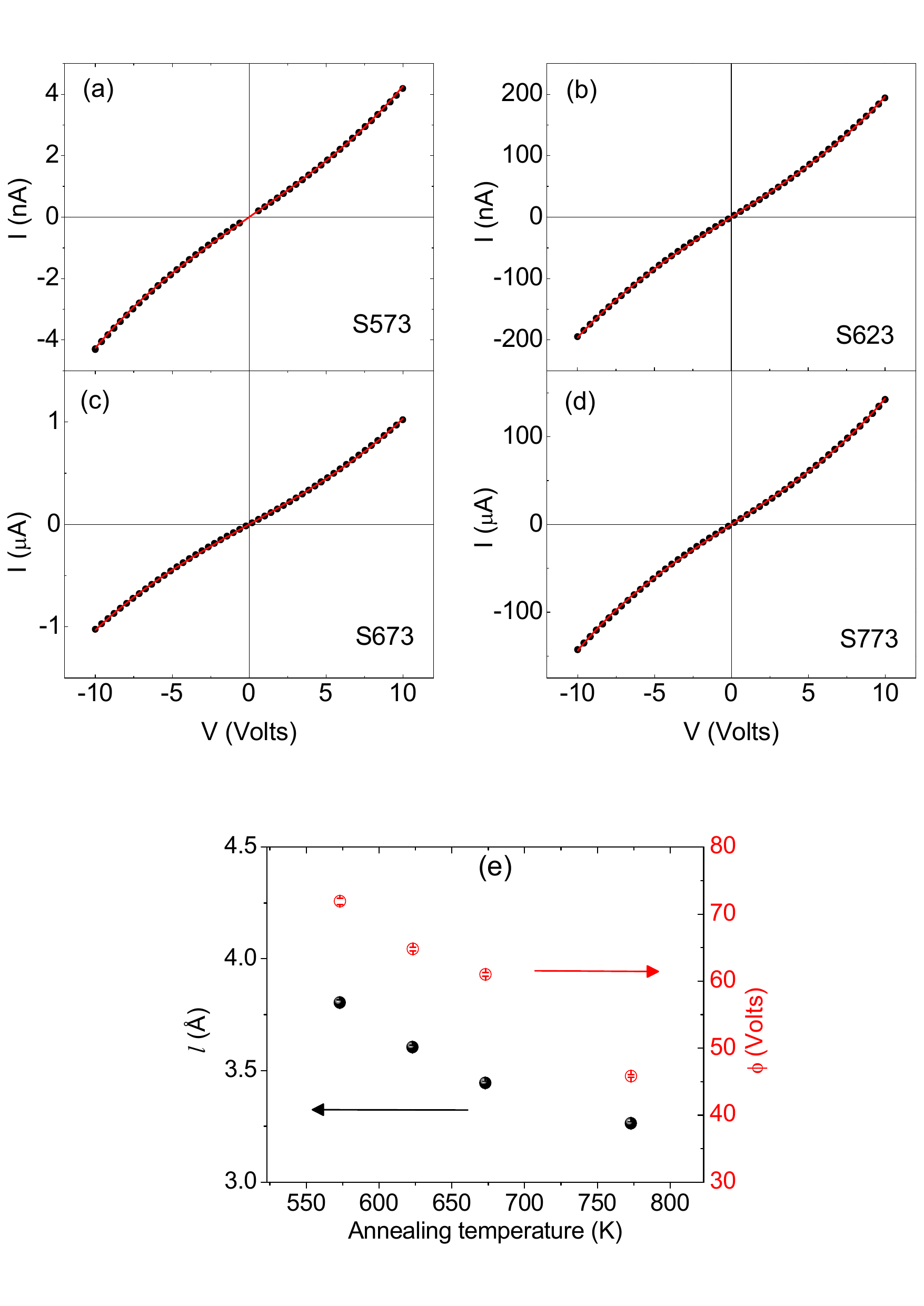}
	\caption{Current-voltage curves measured at 100 K for the samples S573 (a), S623(b), S673(c) and S773(d). The red line is the fit of experimental data to equation (1). The values of $l$ and $\phi$ obtained from the fits are plotted as a function of the annealing temperature in (e).}
	\label{IV}
\end{figure} 

So far, we have shown that heat treatment changes the oxidation state of Fe$_3$O$_4$ nanoparticles self assemblies. Next, we analyze how these characteristics affect the tunnel conduction of the samples. Figure \ref{IV}  shows the current-voltage (I-V) curve of the self-assembled samples measured at 100 K. The non-Ohmic behavior observed in the I-V curve indicates that the electron transport in the devices is related to tunnel conduction. Considering an inelastic tunneling across a rectangular insulating barrier, the barrier height ($\phi$), junction width ($l$) and effective contact area ($a$) can be obtained from the fitting of the I-V curves to the Simmons´s model.\cite{Simmons1963} This model considers a voltage range $0 < V <\phi$:\cite{Vilan2007} 
\begin{equation}
I(V)=6.18\times10^{-10} \frac{a}{l^2}\left\lbrace \left( \phi-\frac{V}{2}\right)exp\left[ -1.025l\left(\phi-\frac{V}{2} \right)  \right] -\left( \phi+\frac{V}{2}\right)exp\left[ -1.025l\left(\phi+\frac{V}{2} \right) \right]  \right\rbrace , 
\end{equation}

where $I$ is expressed in $A$, $a$ in cm$^2$, $\phi$ in V and $l$ in $\mathring{A}$. For the fitting we used the device contact area of each sample considering the electrode length (6-10 mm) and self-assembly film thickness (20 nm $\approx$ two particles thickness) \cite{Fabris2019PRA}. The fittings and the  values of $l$ and $\phi$ obtained from them, are shown in Figure \ref{IV}.
A monotonous decrease of the potential barrier and junction width with the annealing temperature is observed. The value of $\sim$ 3-4 $\mathring{A}$, is considerably  smaller than the estimated carbon gap from the structural measurements, $\sim$ 20 $\mathring{A}$. It is known that Simmon´s model usually gives lower $l$-values than direct structural characterization in magnetic tunnel junction thin films \cite{Buchanan2002}, which is usually attributed to the presence of hot-spots in the barrier, whose density in the present case is expected to increase with the annealing.

\begin{figure}
	\centering
	\includegraphics[width=0.5\textwidth]{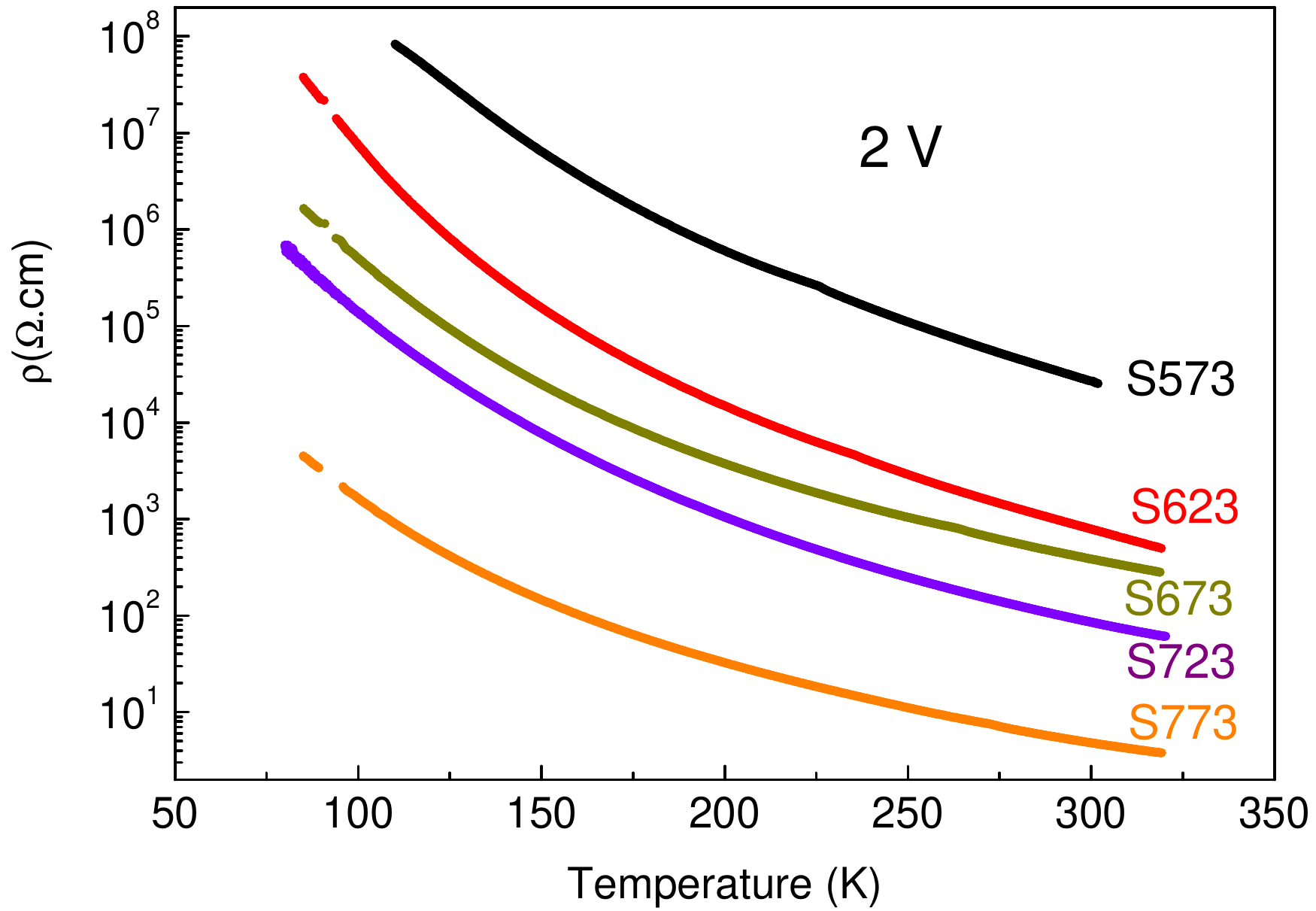}
	\caption{Resistivity as function of temperature measured with an applied voltage of 2 V for all annealed self-assemblies.}%
	\label{RT}
\end{figure}

A monotonous decrease of the resistivity with the temperature is observed for all samples (see the figure \ref{RT}), in agreement with the expected behavior for granular films.\cite{Efros1975,Sheng1973,Zeng2006} Moreover, when the annealing temperature increases, the resistivity systematically decreases, reducing its value more than 4 orders of magnitude for the higher annealing temperature. This behavior is consistent with the reduction of the barrier height and width with the annealing temperature discussed before.

\begin{figure}
	\centering
	\includegraphics[width=\textwidth]{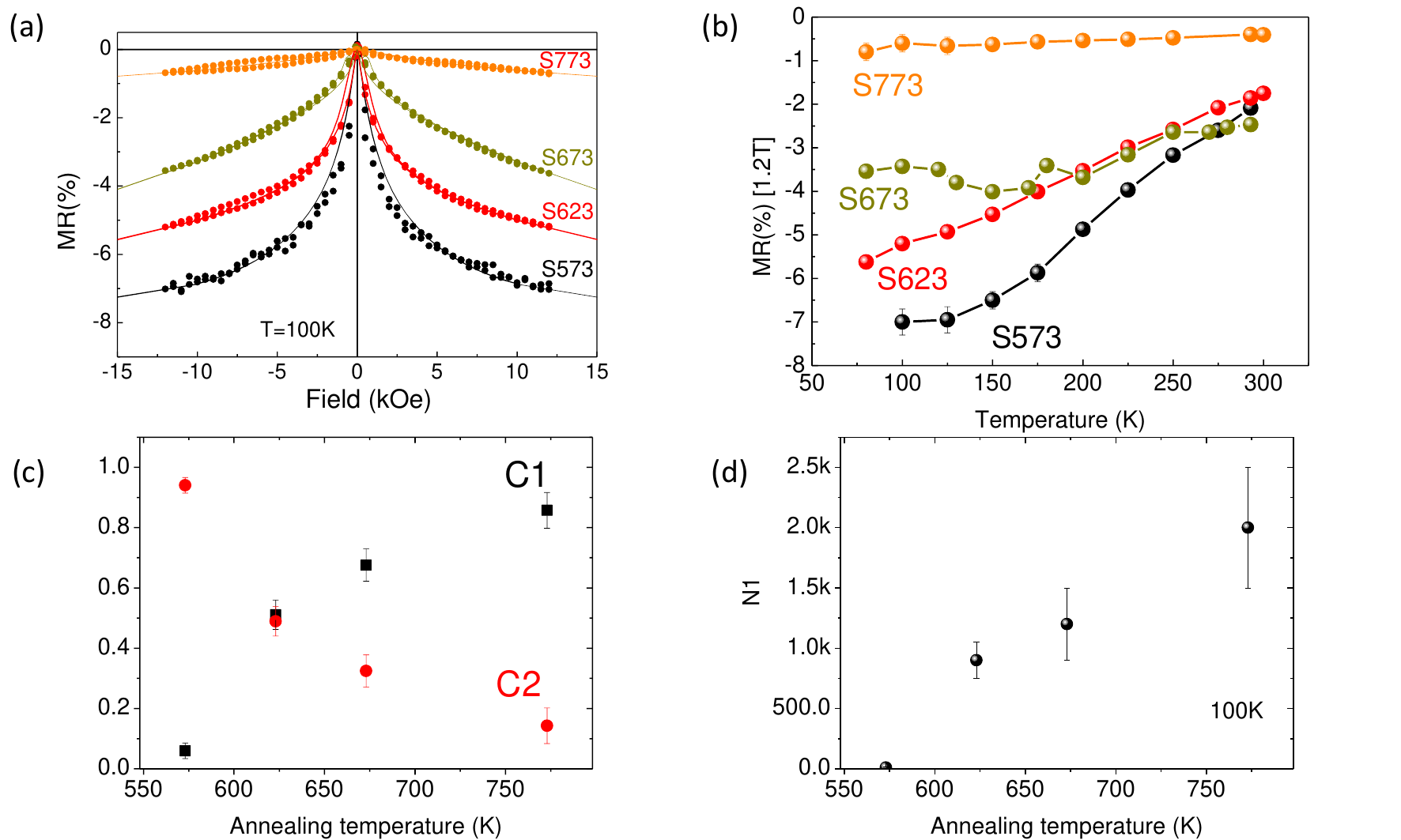}
	\caption{$MR$ at 100 K (a) and the temperature dependence of the $MR$ at 12 kOe (b) for the samples annealed at different temperatures. The solid lines in (a) are the fit of $MR$ with the magnetic loops and scattering through non-collinear surfaces spin described by a Langevin function. The adjusted C1 and C2 parameters are presented in (c) and N1 in (d) as function of the annealing temperature.}
	\label{TMR}
\end{figure}

Figure \ref{TMR}(a) shows the magnetoresistance of the self-assemblies, defined as $MR (H)= 100\times[R(H)+R(0)]/R(0)$, measured at 100 K. The sample S573 present $MR(12 kOe)\sim-7\%$ at 100 K. This value in agreement with previous reports on oleic acid coated Fe$_3$O$_4$ self-assemblies\cite{Jang2007,Wang2009,Taub2009}. For granular material with randomly oriented magnetic easy axis the TMR  is modeled as $TMR=P^2/(1+P^2)$, where $P$ corresponds to the spin polarization at Fermi level\cite{Inoue1996}. Therefore, the decreasing of the magnetoresistance with the annealing temperature observed in Figure \ref{TMR}(a) signals a diminution of the spin polarization with the thermal treatment. This result is consistent with the surface oxidation of the magnetite, which reduces the spin polarization. The presence of magnetic defects at the particles interface may also be the source of spin relaxation which also decreases the $MR$.\cite{Mitani1997}  The thermal dependence of $MR$  is shown in figure \ref{TMR} (b) for all the samples. From this figure it is noticed that the $MR$ of S573 strongly depends on temperature; this dependence diminishes with the annealing temperature, until the $MR$ remains almost temperature independent for the S773. 

In granular systems, the $MR$ can be related to the macroscopic magnetization by the spin-polarized tunneling model: $MR(H,T)\propto T^{-1}[m^2(H,T)-m^2(0,T)]$,\cite{Helman1976,Helman1981} where $m=M/M_S$. The $MR$ curve calculated from this model reproduces well the measurement of sample S573, but it does not have the high field dependence of the other samples, as observed in the figure \ref{TMR-SI} of the Supplementary Material. Notice that the magnetization saturates at high fields, however the $MR$ does not, even at the maximum applied field of 12 kOe. This behavior is usually present in granular systems, where a high-field linear dependence of $MR$ is observed. \cite{Lu2006,Quintela2003,Jang2007} In these systems the interparticle interactions induce spin surface frustration and the spins could be pinned by surface defects which difficult their alignment with the external magnetic field. Therefore, the highly disordered superficial spins dominate the intergranular magnetotransport. \cite{Batlle2002}  
 
Jang et. al.\cite{Jang2007} proposed that the Fe$_3$O$_4$ granular system is formed by particles with spin-aligned cores surrounded by a non-collinear shell. Therefore, the spin dependent scattering has two contributions: one within the nanoparticles, between the spin-aligned core and the noncollinear spins at the shell, which is proportional to the magnetization that can be described by the Langevin function $L(x)$; and a second interparticle contribution between two shells with non-collinear spins proportional to $m^2$. Therefore, $MR$ can be expressed as:
\begin{equation}
MR\propto C_1L(x)+C_2[m^2(H,T)-m^2(0,T)], 
 \end{equation}
where $L(x)= coth(x)-1/x$, $x=N_i\mu_B H/k_BT$, $N_i$ is the number of spins participating in the spin dependent transport and $C_1$ and $C_2$ represent the weighted contribution of the two processes to $MR$. Figure \ref{TMR} shows the fitting of experimental data to this model, with $C_1$, $N_1$ and $C_2$ as fitting parameters, where an excellent agreement with the experimental data is observed. The adjusted parameters are shown in figure \ref{TMR} (c) and (d) as function of the annealing temperature. Notice that the $C_1$ and $N_1$ increase with the annealing temperature while $C_2$ diminishes. This result indicates that in S573 sample the dominant contribution to the magnetoresistance is the intergrain tunneling, which is consistent with the half-metallic nature of the Fe$_3$O$_4$ NPs separated by the residual oleic acid from the synthesis at the assemblies. On the other hand, when the annealing temperature increases, the residual carbon diminishes and the Fe$_3$O$_4$ surface oxidizes, in consequence the interparticle contribution to $MR$ decreases.

\section{Conclusions}

We have shown that vacuum annealing of self-assembled systems of Fe$_3$O$_4$ nanoparticles induces a partial oxidation of the particles, which has important consequences in the magnetic and transport properties: the magnetization decreases and the magnetic anisotropy of the nanoparticles is enhanced, while the resistivity of the sample systematically diminishes with the annealing temperature. More important, the nanoparticles surface oxidation and the increment of defects concentration at the tunnel barrier when the annealing temperature increases, causes a reduction of the spin polarization. In consequence, it is necessary to include in the magnetoresistance an additional term proportional to the magnetization, as the annealing temperature is augmented. Therefore, for nanoparticles self-assembled $MR$ devices, it is important to settle proper thermal treatments, in order to obtain a fine control on the surface composition,  allowing consequently to enhance the tunneling current through the nanoparticles arrange but without degrading the spin polarization due to the surface oxidation or the increasing of defects concentration at the barrier.

\begin{acknowledgments}
 The authors thank to Dr. Guillermo Zampieri for the XPS measurements.The authors also are thankful to Argentine government agency Agencia Nacional de Promoción Científica y Tecnológica (ANPCyT) for the financial support of the work through Grant No. PICT-2019-02059, and Universidad Nacional de Cuyo (UNCuyo) for support through Grants No. 06/C029-T1. F. Fabris acknowledges the São Paulo Research Foundation (FAPESP) for the post-doctoral fellowship with Grant No. 2019/13678-1. The authors also gratefully acknowledge the European Commission for the financial support under the H2020-MSCA-RISE-2016,  SPICOLOST  project Nº 734187 and H2020-MSCA-RISE-2021 ULTIMATE-I project Nº 101007825. F.R. acknowledges financial support of the Ministerio de Economía y Competitividad of Spain (Project No.  MAT2016-80762-R), and Xunta de Galicia (Centro Singular de Investigación de Galicia accreditation 2016-2019) and the European Union (European Regional Development Fund – ERDF).

\end{acknowledgments}

\bibliography{ref}

\newpage

\input {SuplementaryMaterial.tex}

\end{document}

%% file: SuplementaryMaterial.tex
\section{Supplementary Material}

\setcounter{figure}{0}
\renewcommand{\figurename}{Figure}
\renewcommand{\thefigure}{S\arabic{figure}}

\setcounter{table}{0}
\renewcommand{\tablename}{Table}
\renewcommand{\thetable}{S\arabic{figure}}

\begin{figure}[h]
		\centering
		\includegraphics[width=0.8\textwidth]{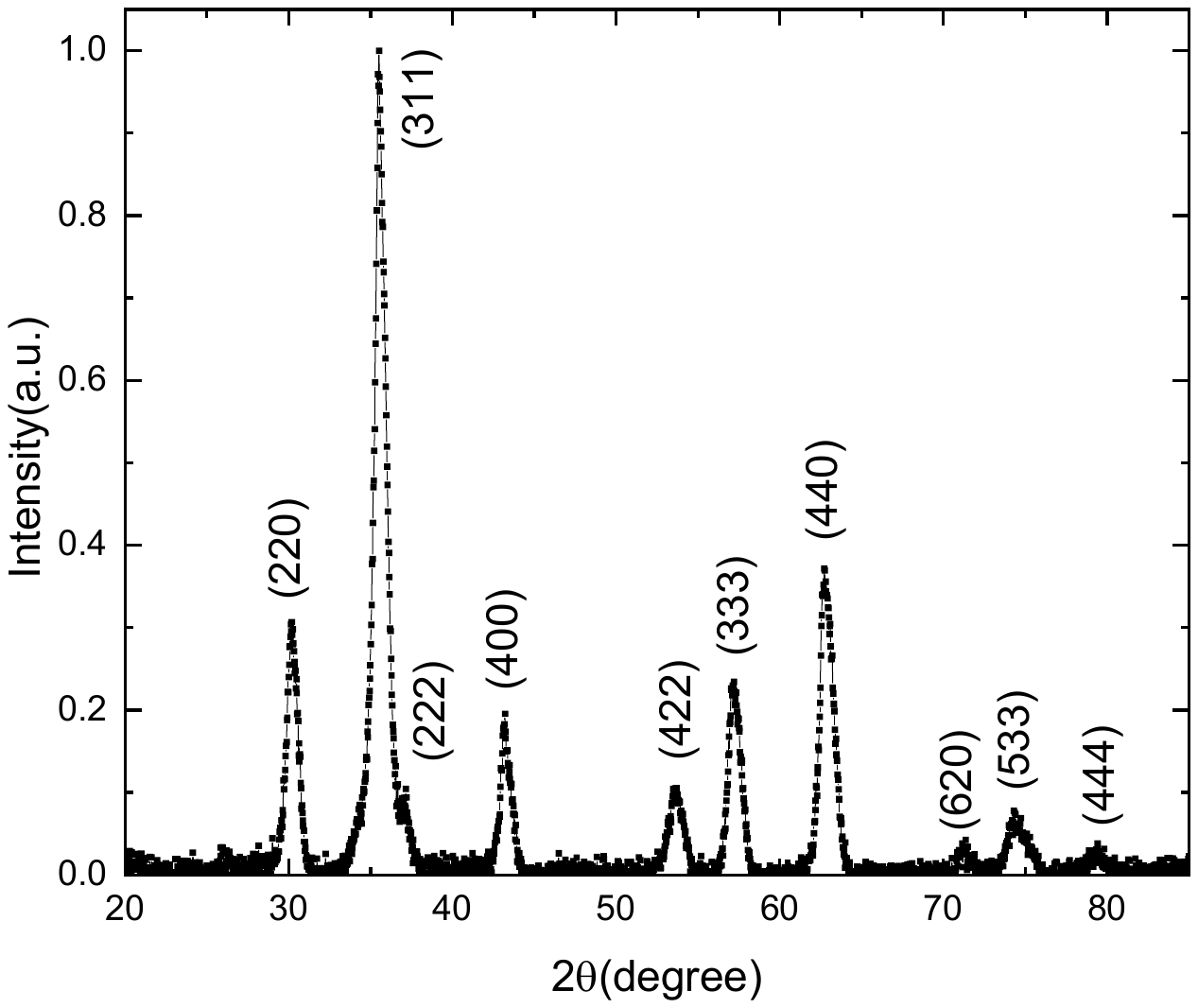}
		\caption{X-ray diffraction pattern of the Fe$_3$O$_4$ as-synthesized nanoparticles where the diffraction peaks are indexed with the Fd3m  space group of the Fe$_3$O$_4$ spinel.}
		\label{XRD}
	\end{figure}

	\begin{figure}[h]
		\centering
		\includegraphics[width=0.8\textwidth]{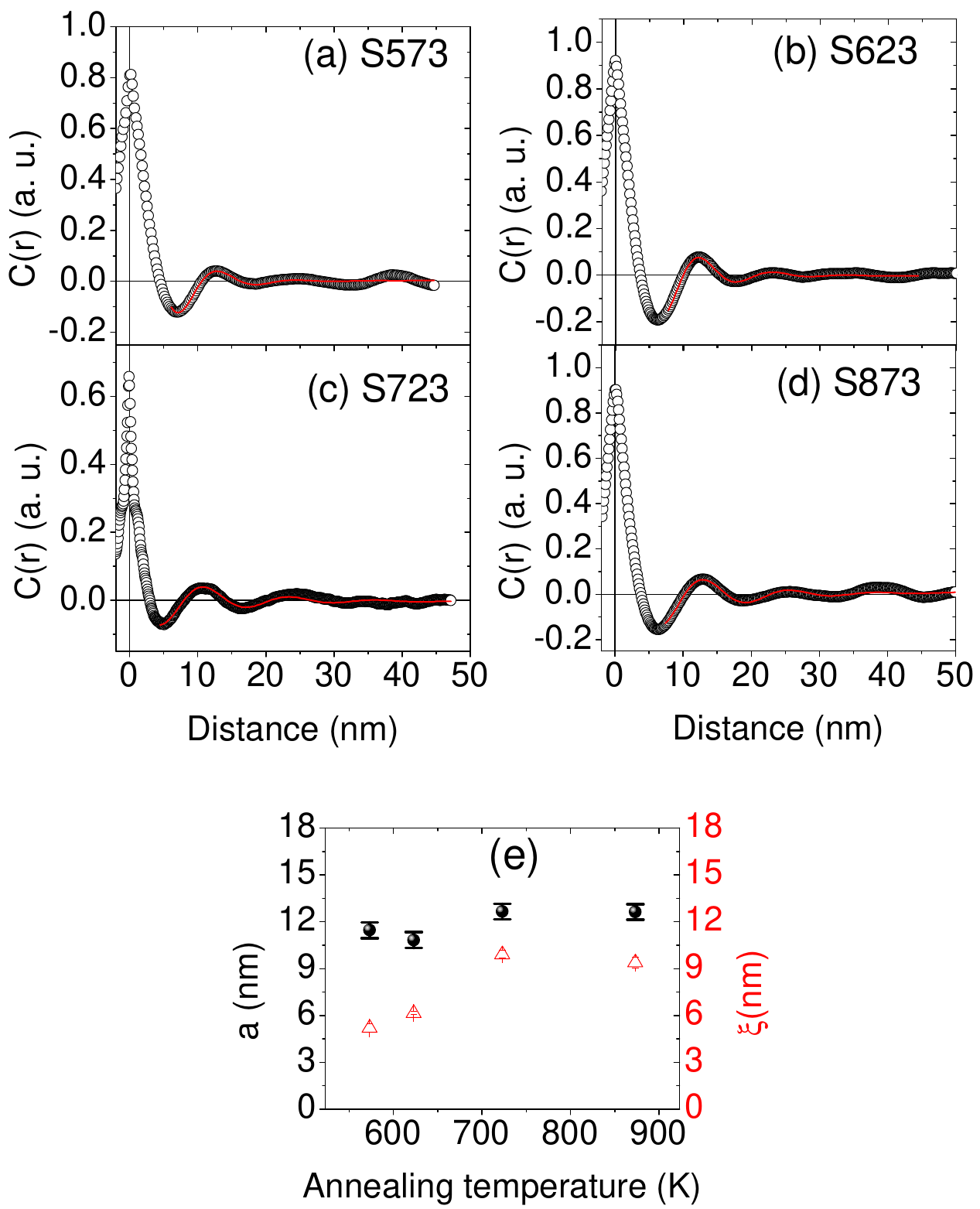}
		\caption{Radial profile of the autocorrelation spectrum   extracted along the red lines presented in the figure \ref{TEM-LM} for the self-assemblies annealed at 573 K (a), 623 K (b), 723 K (c) and 873 K (d). The red line is the fit equation using the equation given in the text. In (e) is shows to adjusted $\xi$  and a parameters as function of the annealing temperature.}
		\label{Correlation}
	\end{figure}

\begin{table}[ht]
	\caption{Adjusted parameters of the XPS spectra at Fe2p region.}
	\resizebox{9cm}{!}{%
		\begin{tabular}{c|c|c|c|c|c}
			\hline
			Sample                 & Component        & B.E. (eV) & $\sigma$ (eV) & Area  & $3\cdot Fe^{2+}/(Fe^{3+}+ Fe^{2+})$ \\ \hline
			\multirow{8}{*}{Fe$_3$O$_4$} & Fe$^{2+}$ $2p_{3/2}$ octa.  & -709.4    & 2.97   & \multirow{2}{*}{12195} & \multirow{8}{*}{0.99}                \\
			& Fe$^{2+}$ $2p_{1/2}$ octa.  & -723      & 3.86   &       &                     \\
			& Fe$^{3+}$ $2p_{3/2}$ octa. & -710.5    & 2.93   & \multirow{2}{*}{13072} &                     \\
			& Fe$^{3+}$ $2p_{1/2}$ octa.  & -724.1    & 3.56   &       &                     \\
			& Fe$^{3+}$ $2p_{3/2}$ tetra. & -712.4    & 4.23   & \multirow{2}{*}{11609} &                     \\
			& Fe$^{3+}$ $2p_{1/2}$ tetra. & -726      & 5.86   &       &                     \\
			& Sat. $2p_{3/2}$        & -718      & 3.5    & \multirow{2}{*}{1629}  &                    \\
			& Sat. $2p_{1/2}$        & -731.6    & 3.79   &       &                     \\ \hline
			\multirow{8}{*}{S573}  &  Fe$^{2+}$ $2p_{3/2}$ octa.  & -709.4    & 2.15   & \multirow{2}{*}{526}   & \multirow{8}{*}{1.11}                \\
			& Fe$^{2+}$ $2p_{1/2}$ octa  & -723      & 4.71   &       &                     \\
			& Fe$^{3+}$ $2p_{3/2}$ octa.  & -710.5    & 1.96   & \multirow{2}{*}{447}   &                     \\
			& Fe$^{3+}$ $2p_{1/2}$ octa.  & -724.1    & 4.5    &       &                     \\
			& Fe$^{3+}$ $2p_{3/2}$ tetra. & -712.4    & 2.49   & \multirow{2}{*}{452}   &                     \\
			& Fe$^{3+}$ $2p_{1/2}$ tetra. & -726      & 9      &       &                     \\
			& Sat. $2p_{3/2}$        & -718      & 4      & \multirow{2}{*}{39}    &                   \\
			& Sat. $2p_{1/2}$         & -731.6    & 2.26   &       &                     \\ \hline
			\multirow{8}{*}{S673}  &  Fe$^{2+}$ $2p_{3/2}$ octa.  & -709.4    & 2.33   & 491   & \multirow{8}{*}{0.68}                \\
			& Fe$^{2+}$ $2p_{1/2}$ octa. & -723      & 3.01   &       &                     \\
			& Fe$^{3+}$ $2p_{3/2}$ octa.  & -710.5    & 2.93   & \multirow{2}{*}{1104}  &                     \\
			& Fe$^{3+}$ $2p_{1/2}$ octa.  & -724.1    & 4.1    &       &                     \\
			& Fe$^{3+}$ $2p_{3/2}$ tetra. & -712.4    & 3.5    & \multirow{2}{*}{575}   &                     \\
			& Fe$^{3+}$ $2p_{1/2}$ tetra. & -726      & 8.06   &       &                     \\
			& Sat. $2p_{3/2}$       & -718      & 4,32   & \multirow{2}{*}{169}   &                    \\
			& Sat. $2p_{1/2}$          & -731.6    & 2.07   &       &                     \\ \hline
			\multirow{8}{*}{S773}  &  Fe$^{2+}$ $2p_{3/2}$ octa.  & -709.4    & 2.2    & \multirow{2}{*}{6716}  & \multirow{8}{*}{0.29}                \\
			& Fe$^{2+}$ $2p_{1/2}$ octa. & -723      & 2.57   &       &                     \\
			& Fe$^{3+}$ $2p_{3/2}$ octa.  & -710.5    & 3.24   & \multirow{2}{*}{33393} &                     \\
			& Fe$^{3+}$ $2p_{1/2}$ octa.  & -724.1    & 3.87   &       &                     \\
			& Fe$^{3+}$ $2p_{3/2}$ tetra. & -712.4    & 4.63   & \multirow{2}{*}{30321} &                     \\
			& Fe$^{3+}$ $2p_{1/2}$ tetra. & -726      & 7.3    &       &                     \\
			& Sat. $2p_{3/2}$        & -718      & 4.89   & \multirow{2}{*}{14958} &                    \\
			& Sat. $2p_{1/2}$         & -731.6    & 8.45   &       &                     \\ \hline
			\multirow{8}{*}{S873}  & F Fe$^{2+}$ $2p_{3/2}$ octa.  & -709.4    & 1.38   & \multirow{2}{*}{1093}  & \multirow{8}{*}{0.13}                \\
			& Fe$^{2+}$ $2p_{1/2}$ octa. & -723      & 5.69   &       &                     \\
			& Fe$^{3+}$ $2p_{3/2}$ octa.  & -710.5    & 2.44   & \multirow{2}{*}{15097} &                     \\
			& Fe$^{3+}$ $2p_{1/2}$ octa.  & -724.1    & 3.22   &       &                     \\
			& Fe$^{3+}$ $2p_{3/2}$ tetra. & -712.4    & 3.42   & \multirow{2}{*}{9320}  &                     \\
			& Fe$^{3+}$ $2p_{1/2}$ tetra. & -726      & 9      &       &                     \\
			& Sat. $2p_{3/2}$        & -719.4    & 2.73   & \multirow{2}{*}{1927}  &                    \\
			& Sat. $2p_{1/2}$        & -733      & 2.46   &       &                     \\ \hline
		\end{tabular}%
	}
	
	\label{XPS-results}
\end{table}
	
\begin{figure}
	\centering
	\includegraphics[width=0.8\textwidth]{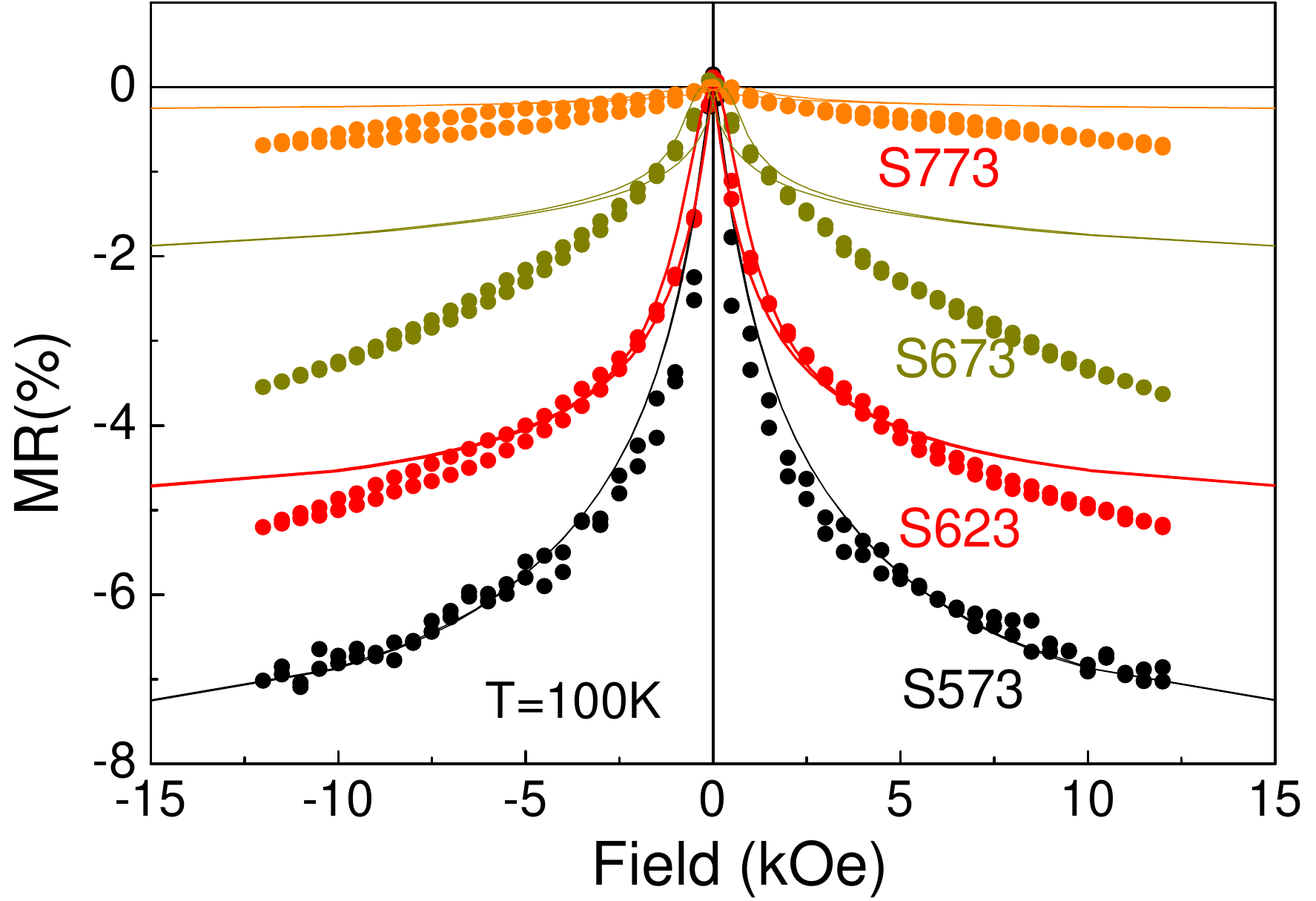}
	\caption{Comparison of the TMR (solid dots) with the $-M^2(H)$ (solid lines) measured at 100 K.}
	\label{TMR-SI}
\end{figure}